\def\etal{{\em et al.~}}

\def\figdir{ }

\documentstyle[11pt,aaspp]{article}
\eqsecnum

\slugcomment{To appear in the {\em The Astrophysical Journal Letters}, 1 October 1997}

\begin{document}

\title{On Nulling Interferometers and the Line-Emitting Regions of AGNs}

\author{G. Mark Voit}
\affil{STScI, 3700 San Martin Drive, 
       Baltimore, MD 21218}

\setcounter{footnote}{0}

\begin{abstract}
The nulling interferometers proposed to study planets around 
other stars are generally well suited for studying 
small-scale structures surrounding other bright pointlike objects
such as the nuclei of active galaxies. Conventional interferometric 
techniques will produce useful maps of optical/IR line 
and continuum emission within active galaxies on scales 
of 10 milliarcseconds (mas), but similar studies of 
broad-line regions will require baselines 
longer than those currently envisaged.  
Nevertheless, nulling interferometers currently under development
will be able to constrain quasar velocity fields on $\sim 1$~mas 
scales, as long as they are equipped with spectrographs capable
of resolving lines several hundred km~s$^{-1}$ wide.  This 
{\em Letter} describes how analyses of line emission leaking
through the edges of the null in such an instrument can reveal
the size, shape, and velocity field of nebular gas
on the outskirts of a quasar broad-line region.  If this technique
proves effective, it could potentially be used to measure the
mass function of quasar black holes throughout the universe.
\end{abstract}

\keywords{galaxies: active --- quasars: emission lines ---
techniques: interferometric}

\section{Introduction}

Nulling interferometry, a technique inspired by the desire to
study extrasolar planets, appears ready to blossom into a practical 
astronomical tool in the coming decades 
(Allen, Peterson, \& Shao 1997; Angel \& Woolf 1997).  
The most basic nulling interferometer would consist of two 
identical telescopes and a beam combiner that inverts 
the phase of one telescope's signal before adding it 
to the other (Bracewell 1978; Bracewell \& Macphie 1979;
Shao \& Colavita 1992).  
If active delay lines can equalize all the pathlengths 
in the optical system, the signal from a pointlike object 
in the symmetry plane between the two telescopes will cancel 
at all wavelengths. 

The signal cancelling capability of nulling interferometers 
makes them ideal for observing planets around nearby stars, 
as well as other sources less than an arcsecond from a bright
object. These devices will be excellent for studying the  
narrow-line regions of active galaxies. Maps of the
$2 - 10 \mu$m emission around active galactic nuclei 
(AGNs) will help resolve questions ranging from the role
of starbursts in AGNs to the anisotropy of their UV emission 
(Miley 1997; Voit 1997a,b; Ward 1997).  Even without nulling,
ground-based continuum observations will easily resolve
the thermal emission from 300~K material, because active 
optical correction of large telescopes is relatively simple 
at 10~$\mu$m.

Broad-line regions of AGNs present a greater challenge.  
Their typical sizes, a few milliarcseconds or less,
are somewhat smaller than the smallest beamsizes currently
proposed for optical/IR interferometers. The Keck and Very Large 
Telescope (VLT) interferometers, with baselines $\sim 100$~m, 
will have beamsizes $\sim 4$~mas at 2~$\mu$m.  The Space 
Interferometry Mission (SIM), slated to operate in the optical 
band, envisions a baseline of 10~m, giving a minimum beamsize 
$\sim 10$~mas.  Although SIM's nominal beamsize is somewhat
larger, its position above the Earth's distorting atmosphere 
makes it especially useful for nulling observations, which require 
precise management of the total optical path from the source 
through the instrument.  

Despite these beamsizes, nulling interferometers might still 
yield unique information about the broad-line regions of AGNs, 
if we tailor the instruments properly and keep our goals modest. 
This {\em Letter} proposes a few strategies for 
observing the broad-line regions of AGNs with nulling
interferometers, emphasizing the benefits of coupling these
instruments with moderate-resolution spectrographs.

\section{Possibilities}

Nulling interferometers possess useful coronagraphic 
capabilities on scales smaller than their beamsizes.
Each pointing of such an instrument convolves 
its two-dimensional transmission function with 
the projection of the target onto the plane of the sky.
In principle, we can extract information about the 
source on scales smaller than the nominal resolution 
element as long as we know the interferometer's transmission 
characteristics accurately. 

Active galactic nuclei are ideal for such experiments because 
their spectra depend on projected distance from the nucleus.
For example, optical long-slit spectroscopy of M87 with the Faint 
Object Camera on HST has revealed that its emission lines
obey a nearly Keplerian rotation law to within 200 mas of
its nucleus, where the projected rotational velocity is
$\sim 600 \, {\rm km \, s^{-1}}$ (Marconi \etal 1997; 
Macchetto \etal 1997).  In the coming years, the Space
Telescope Imaging Spectrograph will yield rotation curves 
on similar scales for a number of other nearby galactic 
nuclei.

Spectroscopic instruments with significantly higher spatial 
resolution will be needed to probe the velocity fields within 
distant quasars.  Reverberation mapping experiments can provide
information about AGNs on tiny scales of light days to months 
(e.g. Korista \etal 1996), regardless of the source's distance, 
but the interpretation of this information can be ambiguous 
(Done \& Krolik 1996; Chiang \& Murray 1996).  In addition, 
the long variability timescales of distant quasars, owing to 
both their larger sizes and cosmological time dilation, render 
reverberation mapping of high-$z$ objects impractical.  
Nulling interferometry, combined with spectroscopy, might be 
our best hope for weighing high-$z$ black holes within
the next decade.

\subsection{Combining Spectroscopy with Nulling}

Consider a nulling interferometer with a baseline $B$ 
operating at a wavelength $\lambda$, and let $\delta$ be 
the angular distance between a given point on the 
sky and the locus of the symmetry plane. The transmission 
function of a such two-element interferometer is
$T(\delta) = 1 - \cos \delta / \delta_0$, where 
$\delta_0 \equiv \lambda / 2 \pi B$. 
For an optical instrument like SIM with a 10~m baseline, 
$\delta_0 \approx 1.6 \, {\rm mas}$ and $T(1 \, {\rm mas}) 
\approx 0.2$, whereas the transmission of a point source 
directly on the null is $\sim 10^{-4}$ (Allen \etal 1997). 

Simplistically we can picture an AGN as a pointlike 
continuum source surrounded by an extended line-emitting 
region.  If we could aim a nulling interferometer directly 
at such an idealized AGN, the continuum signal would cancel, 
but off-axis line emission would leak through the interferometer. 
A spectrograph mounted on the interferometer would record 
only an emission line with a strength depending on the size 
and shape of the line-emitting region.  

A single pointing of such a device at an AGN would thus 
tell us immediately the extent of the line-emitting 
region relative to the interferometer's beam. Comparing the 
total flux of an emission line ($F^{\rm tot}$) measured 
through a single-element system with that measured through 
a nulling interferometer ($F^{\rm null}$) places a lower 
limit on the region's size:
$\delta > \delta_0 \cos^{-1} [1 - (F^{\rm null} / F^{\rm tot})]$.
The continuum level would respond to the extended
background from stars in the host galaxy and
perhaps some scattered continuum light from the nucleus.

Modulations in the line flux transmitted through a rotating 
interferometer would echo the symmetry of the line emitting region.
A linearly extended region, for example, would produce a line-flux
signal that cycles from minimum to maximum and back 
twice during one complete rotation of the instrument.
If the so-called ``unified models'' of AGNs are correct, the
degree of azimuthal structure detected in this way should 
correlate with the inclination angle of the obscuring torus.
Such information could also reveal whether the line emission
comes from a disk or torus perpendicular to a jet or
from entrained gas parallel to the jet itself.

\subsection{Probing Milliarcsecond Velocity Structure}

Assuming that a black hole of mass $(10^9 \, M_\odot) M_9$ drives
the activity within any given active galaxy or quasar, we can
easily compute the angular size $\theta$ of a $(1000 \, {\rm km 
\, s^{-1}}) v_{1000}$ Keplerian orbit around the central engine.  
Figure~1 gives the angular sizes of such orbits as functions 
of redshift $z$ and density parameter $\Omega$ for various 
values of the parameter combination $M_9 h_{75} v_{1000}^{-2}$, 
where $h_{75}$ is the Hubble constant in units of $75 \, 
{\rm km \, s^{-1} \, Mpc^{-1}}$.  Because the Eddington 
luminosity of a quasar is $(\sim 10^{47} \, {\rm erg 
\, s^{-1}}) M_9$, the velocity fields around the most luminous 
quasars should exceed $1000 \, {\rm km \, s^{-1}}$
at projected distances of 1~mas, no matter what the redshift.
A nulling interferometer attached to a spectrograph of moderate
resolving power ($R \equiv \lambda / \Delta \lambda \gtrsim 300$)
can therefore constrain the velocity field at the outskirts 
of the broad-line region around any sufficiently luminous quasar.
   
If the velocities of broad-line emitting clouds are completely 
random or if the line-emitting gas flows out in a wind that
reaches its terminal velocity within the core of the 
null, we would expect the widths of the emission lines 
seen through an interferometer to match those seen through
a single-element telescope.  On the other hand, if the 
line profile at a given projected radius depends on the 
gravitational potential of the nucleus, the width 
of the transmitted line should depend on the baseline 
of the interferometer. The following highly simplified 
model illustrates how the results of such an experiment 
might look. 

Suppose that the target hosts a circularly symmetric 
emission-line region with a projected angular radius 
$\theta_{\rm BLR}$ at which the Keplerian velocity is 
$1000 \, {\rm km \, s^{-1}}$, and consider two possible 
velocity laws.  In the Keplerian case, the line profile 
at a given angular radius $\theta$ is 
\begin{equation}
  f_{\rm Kep}(v,\theta) \propto \exp \left[ - \left( 
                  \frac {v} {1000 \, {\rm km \, s^{-1}}} \right)^2
                  \frac {\theta} {\theta_{\rm BLR}}  \right] \; .
\end{equation}
And in the case of a uniform velocity law,
\begin{equation}
  f_{\rm uni}(v) \propto \int_0^{\theta_{\rm BLR}} 
			 f_{\rm Kep}(v,\theta) \,
                         \theta \, d \theta \; ,
\end{equation}
so that at low spatial resolution the line profiles are
indistingishable. 

The observed line profiles resulting from these velocity laws 
look markedly different after the light passes through a nulling 
interferometer. One effective way to visualize differences between 
the overall line profile ($F_\nu^{\rm tot}$) and the nulled line 
profile ($F_\nu^{\rm null}$) in each case is to divide 
$F_\nu^{\rm null}$ by $F_\nu^{\rm tot}$.  Whenever the velocity 
law is uniform, this quotient will be a horizontal line. 
A declining velocity law, such as $f_{\rm Kep}$, will show 
a peaked quotient, and the quotient for an accelerating flow 
should rise away from a minimum at 0~km~s$^{-1}$.

Figure~2 illustrates the quotient $F_\nu^{\rm null} / 
F_\nu^{\rm tot}$ for a few selected parameter sets.  The
surface brightness $I(\theta)$ of the emitting region 
decreases as $\theta^{-2}$, so that each logarithmic interval
in $\theta$ contributes a similar proportion of flux.  
We truncate the outer radius of the emitting region at
$\theta_{\rm BLR} = \delta_0$ and $2 \delta_0$, equivalent
to observing the same region at two different baselines, one 
double the other.  In order to avoid a divergence in flux,
we truncate the emitting region at an inner radius of $0.01 \, 
\theta_{\rm BLR}$. Solid lines show the resulting quotient 
profiles for $f_{\rm Kep}(v,\theta)$, and dashed lines show 
the equivalent profiles for $f_{\rm uni}(v)$. The dotted lines
illustrate error estimates to be discussed in the following 
section.

The flux quotients in Figure~2 clearly reveal several basic
characteristics of the underlying models.  Models with 
Keplerian velocity laws produce pronounced peaks, indicating
a decline in velocity dispersion with radius.  When $\delta_0 
= \theta_{\rm BLR} / 2$, we find $F^{\rm null} / F^{\rm tot} 
\approx 0.19$, indicating that the transmitted emission comes
from a characteristic radius $\gtrsim 0.31 \, \theta_{\rm BLR}$.  
When $\delta_0 = \theta_{\rm BLR}$, we find $F^{\rm null} / 
F^{\rm tot} \approx 0.05$, indicating a characteristic 
transmission radius $\gtrsim 0.32 \, \theta_{\rm BLR}$.
In both cases, $F^{\rm null}_\nu$ is nearly proportional 
to $\exp[-(v/1200 \, {\rm km \, s^{-1}})]$. Naively associating
this velocity scale with the characteristic radius $0.32 \,
\theta_{\rm BLR}$ would yield an estimated central mass about
half the true value.  Invoking reasonable assumptions about 
$I(\theta)$, which will in general be unknown, would result 
in a more accurate but somewhat model dependent mass estimate.

\section{Practical Matters}

Techniques such as the one described here could potentially be 
used to weigh black holes within high-redshift AGNs if SIM or 
other nulling interferometers were equipped with spectrographs 
capable of resolving broad emission lines.  If the velocity fields
of broad-line regions turn out to reflect their gravitational 
potentials, as appears likely from HST observations of nearby
AGNs (e.g. Marconi \etal 1997, Macchetto \etal 1997), 
interferometric measurements of these lines might 
tell us how close the luminosities of the central objects are 
to their Eddington limits, perhaps revealing how the black hole 
mass function in quasars changes with time.  To be practical,
these observations will have to overcome photon noise limitations
of various sorts.

The current plan for SIM calls for 7 siderostats, each 30~cm in 
diameter, only two of which can be used simultaneously for 
nulling interferometery (Allen \etal 1997).  A quasar with 
an AB magnitude of 17.5 at a particular wavelength $\nu$ 
produces a continuum photon flux at Earth of $F_\nu / h =
0.56 \, {\rm phot \, cm^{-2} \, s^{-1}}$.  Let the equivalent
width of a line like C~IV $\lambda$1550 in the spectrum of
such a quasar be $W_\lambda / \lambda = 0.1 \, W_{0.1}$.
The total number of line photons impinging on two SIM siderostats 
in a $(1 \, {\rm day})t_{\rm day}$ exposure would then be 
$(6.8 \times 10^6) W_{0.1} t_{\rm day}$.  Obtaining $10^5$~counts 
in the spectrograph detector should be quite achievable, even with
very modest assumptions about the instrumental efficiency.

The dotted lines in Figure~2 illustrate the 3$\sigma$ error
boundaries resulting from photon noise in $F^{\rm null}_\nu 
/ F^{\rm tot}_\nu$, assuming that each exposure is long enough 
to yield $10^5$~counts in the absence of nulling.  The nulled 
spectra therefore comprise $5 \times 10^3$ counts and $2 
\times 10^4$ counts for $\delta_0 = \theta_{\rm BLR}$ and 
$2 \theta_{\rm BLR}$, respectively.  At a resolution $R = 300$, 
the spectrograph divides the photons into 1000~km~s$^{-1}$ bins, 
seven of which cover the interesting part of the line.  
The precision of these seven points would be ample for determining 
the shape, amplitude, and velocity width of the flux quotient.  

Ground-based interferometers with much more collecting area (i.e. 
Keck, VLTI) could conceivably be used to carry out similiar 
experiments on infrared lines from much fainter sources if there is a
nearby calibrating source in the field bright enough for 
active fringe tracking.  Bright quasars might be sufficiently 
luminous to serve as their own fringe-tracking sources; 
a $K$ magnitude of 16 would probably suffice (von der Luhe 1997).
Because the $\nu F_\nu$ spectrum of an AGN is generally constant 
to within a factor of a few, the photon flux within an interval of
order $\nu$ is roughly proportional to $\nu^{-1}$, making fringe
tracking easier at longer wavelengths.

Projected space-based interferometers optimized for planet-finding 
(e.g. Angel \& Woolf 1997) have larger mirrors than SIM
and baselines longer by an order of magnitude.  Such 
instruments could potentially study regions light days in size
in the nearest active galaxies, complementing reverberation-mapping
investigations, if they were engineered to work in the optical/UV 
as well as the IR.  UV optimization would be unnecessary for 
studying high-$z$ quasars because their strongest lines shift 
into the optical and intergalactic hydrogen obscures their UV light.  

To compensate for the vibrational modes of such a large space 
structure the instrumentation will have to track source fringes 
with great accuracy (Woolf, Angel, \& Burge 1997).
The nulling precision required for planet finding ($\sim 10^{-3}$
in amplitude and phase, $\sim 10^{-6}$ in flux) is much stricter 
than that required for the types of observations described here 
($\sim 10^{-1}$ in amplitude and phase, $\sim 10^{-2}$ in flux),
but the typical brightnesses of quasars are $\lesssim 10^{-5}$ 
times those of nearby solar-type stars.  Thus, interferometric 
studies of distant quasars would benefit from mirrors as large
as possible -- at least large enough to capture hundreds of photons
within one vibrational period of the fastest significant mode.

\section{Summary}

This {\em Letter} has suggested how nulling interferometers
might be used to probe the line emitting regions of AGNs on
milliarcsecond scales. If instruments such as SIM are equipped
with spectrographs with sufficient resolution, we will be able
to constrain the velocity fields of line-emitting gas around 
luminous quasars throughout the universe.  If these velocity fields 
are Keplerian, we might be able to weigh black holes at high 
redshifts, perhaps learning how their mass function evolves 
with time.

\vspace*{2.0em}

I thank Harley Thronson for encouraging me to consider the 
interesting extragalactic science that might be done with 
nulling interferometers.  I also thank the referee, Nick
Woolf, for his helpful comments.

\pagebreak

\pagebreak

\begin{figure}
\plotone{\figdir 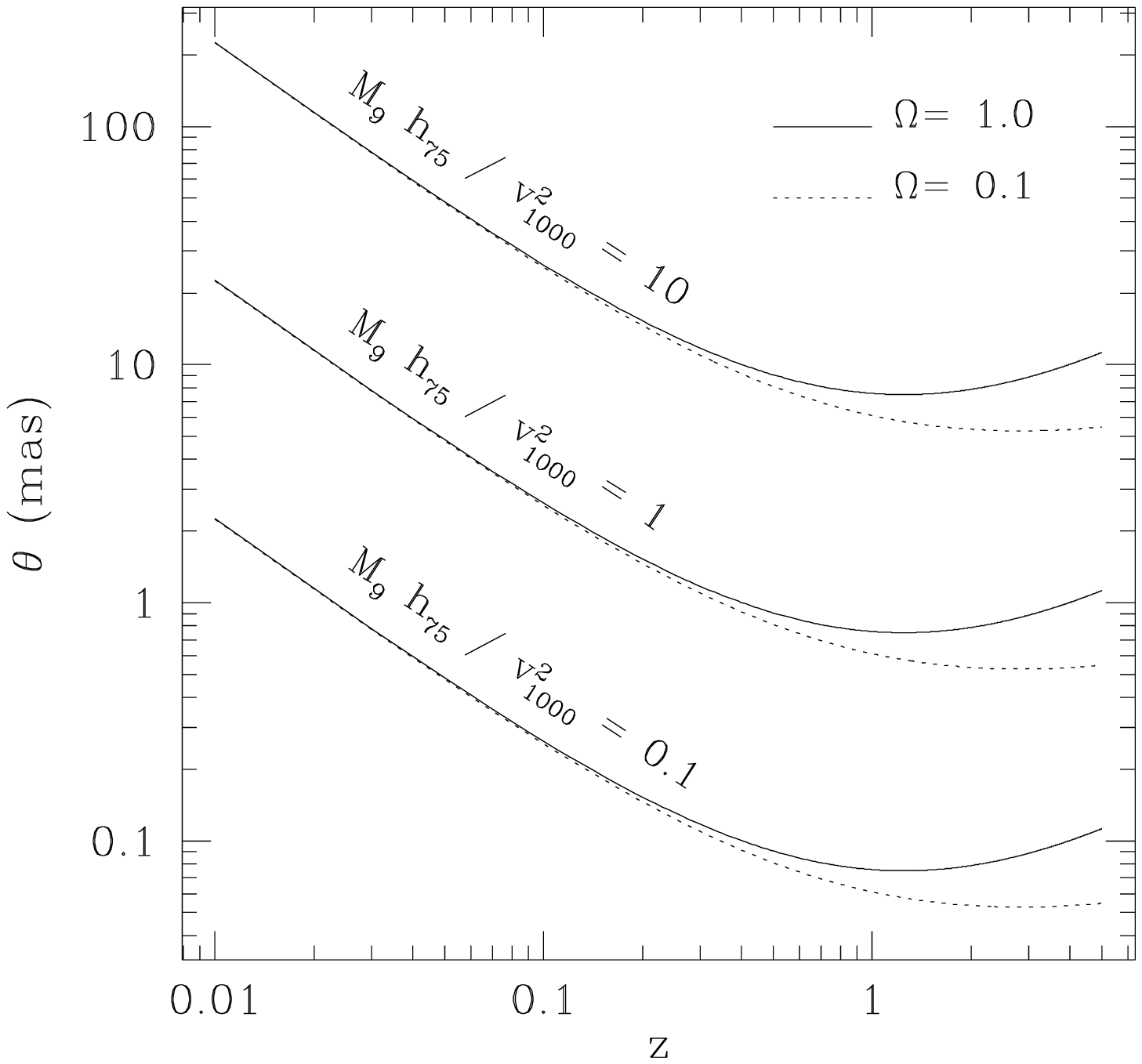}
\caption{Angular sizes of Keplerian orbits. The lines show the angular
sizes of circular orbits at speeds $v_{1000} (1000 \, {\rm km \, s^{-1}})$
around objects of mass $M_9 (10^9 \, M_\odot)$, given a Hubble constant
$h_{75} (75 \, {\rm km \, s^{-1}})$.
}
\label{deltakep}
\end{figure}

\begin{figure}
\plotone{\figdir 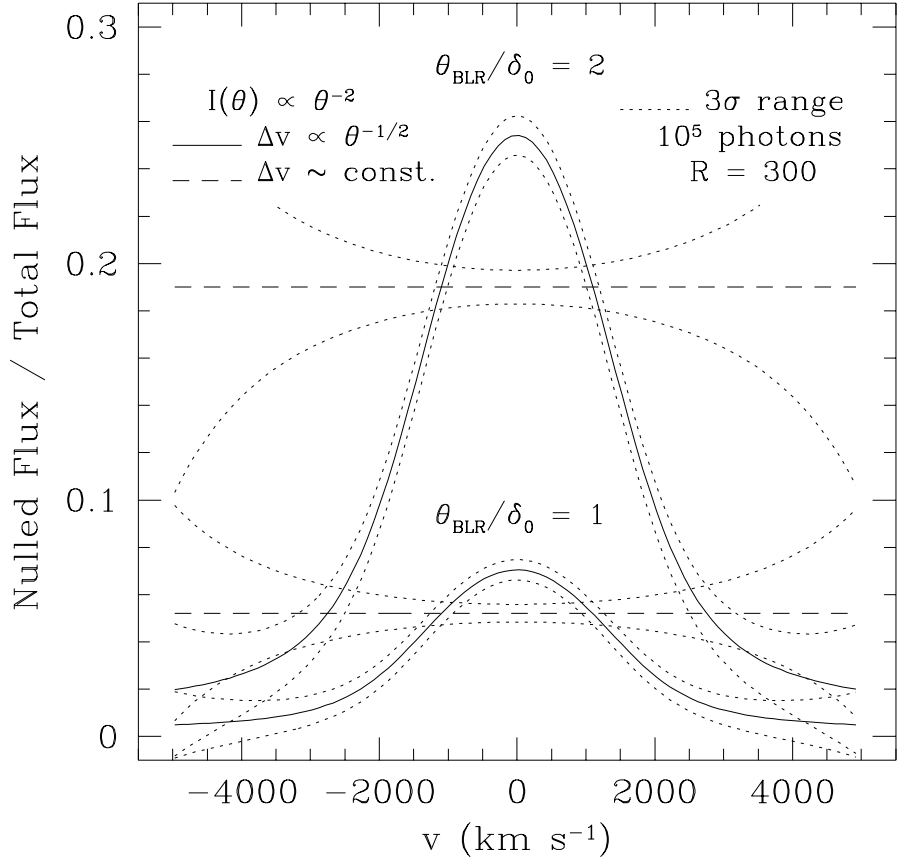}
\caption{Relative transmission through a nulling interferometer.
Solid lines illustrate the quotient spectrum $F_\nu^{\rm null}/
F_\nu^{\rm tot}$ of an emission line region whose line profile
at each radius is a Gaussian with a dispersion $\Delta v \propto 
\theta^{-1/2}$.  The surface brightness $I$ is taken
to decrease with angular radius $\theta$ like $\theta^{-2}$.
The dashed lines show the quotient spectrum of a similar region
with constant line widths.  Dotted lines give the 3$\sigma$ limits
on photon shot noise at spectral resolution $R = 300$ when the 
integration time, assumed the same for all spectra, yields $10^5$ 
counts in the integrated flux $F^{\rm tot}$.
}
\label{nullspecs}
\end{figure}
\end{document}